# Pulse-Burst Excitation Reveals Time–Dose Reciprocity Breakdown in Mixed-Halide Perovskites

Alexandr Marunchenko[1], Shivam Singh [2,3], Daniel Lizotte[1], Bhaskar De[1], Yana Vaynzof [2,3], Ivan G. Scheblykin[1*]

[1] Chemical Physics and NanoLund, Lund University, P.O. Box 124, 22100 Lund, Sweden

[2] Chair for Emerging Electronic Technologies, Technical University of Dresden, Nöthnitzer Str. 61, 01187 Dresden, Germany

[3] Leibniz Institute for Solid State and Materials Research Dresden, Helmholtzstraße 20, 01069 Dresden, Germany

*) ivan.scheblykin@chem.lu.se

## Abstract

Time–dose reciprocity, commonly associated with the Bunsen–Roscoe law, states that the response of a photosensitive system depends only on the total exposure dose, regardless of how that energy is delivered over time. Light-sensitive processes in mixed-halide perovskites, such as photoinduced halide segregation, often exhibit threshold-like behavior that may violate this principle and enable material-state control by photon timing. We test this using pulse-burst excitation, which introduces an additional temporal control dimension beyond conventional parameters such as pulse fluence, repetition rate, and average power. By redistributing the same photon dose over microsecond-to-millisecond timescales, we create distinct nonequilibrium excitation conditions and show that mixed-halide perovskites can evolve into different metastable states, revealing a breakdown of time–dose reciprocity in the combined processes of halide segregation and remixing. This additional temporal degree of freedom not only enables control of the material state but also provides a new experimental framework for disentangling the competing processes underlying photoinduced halide redistribution. Our findings establish photon timing as a control parameter for perovskite photochemistry and open additional opportunities for optical memory and neuromorphic photonic applications.

The simplest model of a photosensitive material is a photon counter, where absorbing the same photon dose leads to the same final material state – the operation regime one expects from a good photodetector. This underlies the Bunsen–Roscoe time–dose reciprocity law[1]. Photographic films are a classic example of how this law fails [2]. In photography, varying exposure times over a wide range while keeping the total photon dose constant results in different degrees of darkening after development, a phenomenon also known as the Schwarzschild effect[2,3], discovered in the context of astronomical imaging.

The microscopic reason for the breakdown of the reciprocity law in photography is the nature of the photoinduced nucleation process. In the Gurney–Mott picture, light creates electronic carriers in silver-halide crystals, which later, when trapped, help to reduce mobile silver ions and form small metallic silver clusters[4]. However, only when these clusters reach a sufficient size can they become developable latent-image centers.[5,6] This threshold, together with the coexistence of multiple light-induced electronic, ionic, and photochemical processes, leads to the breakdown of reciprocity.

Mixed-halide perovskites are modern semiconductors that are, in some ways, analogs to photographic films. Under illumination, initially homogeneously distributed halides, such as iodide and bromide, can undergo redistribution, eventually forming local iodide-rich nano-domains with a small bandgap embedded in a bromide-rich surrounding with a higher bandgap [7–10]. Optically, this is observed as a redshifted photoluminescence (PL) spectrum [7,8]. But in the absence of light, or under specific illumination conditions, the material can partially [7,11–13] or fully [14,15] return to its initial state, re-establishing homogeneous mixing of halides. This photoinduced segregation–mixing equilibrium was shown to depend on light intensity [16,17], fluence [18], repetition rate [18,19,12], temperature[20,13,21], material composition[22,11,19], and defect landscape[23–26,19]. Importantly, the light-induced segregation has a characteristic intensity threshold [16,27], establishing mixed-halide perovskites as nonlinear photosensitive media rather than simple photon counters.

Illumination-dependent studies of various optical and electronic properties of mixed-halide perovskites have been conducted to elucidate the mechanisms underlying their light sensitivity and to explore possibilities for their control by light. These studies investigated the regimes in which segregation appears and disappears, but in most cases, the photon dose delivered to the mixed-halide perovskites was not held constant [14,15,18,28] because changing repetition rate or pulse fluence independently naturally changes the photon dose. At the same time, changing them together places the material under very different excitation conditions[18,19], due to nonlinear fluence-dependent charge recombination processes in perovskites[29]. Therefore, the role of photon timing has remained difficult to isolate so far. Yet, a clear motivation comes from medicine [30–33], biology [1,34], and photosensitive materials [35,36], which provide examples of how photon timing determines the response and performance of photosensitive systems. We thus find it important to test time–dose reciprocity in mixed-halide perovskites under tailored photon timing.

In this work, we realize light exposure under tailored photon timing by applying pulse-burst excitation. Identical laser pulses are grouped into periodic packets (bursts) while maintaining the same pulse fluence and total photon dose. We show that varying the pulse-burst parameters drives mixed-halide perovskites into distinctly different states under equivalent photon-dose conditions, as revealed by their steady-state PL spectra. Our measurements directly reveal long-lived populations of trapped charges and associated photodoping that persist well beyond the prompt carrier lifetime, providing a memory of preceding excitation. These hidden photoexcitations are therefore likely to contribute to the sensitivity to photon timing. Together, these results reveal a breakdown of time–dose reciprocity and establish photon timing as an additional control dimension for structural phases in mixed halide perovskites. Finally, we demonstrate reversible switching between these states over more than a thousand segregation–mixing cycles.

## Results

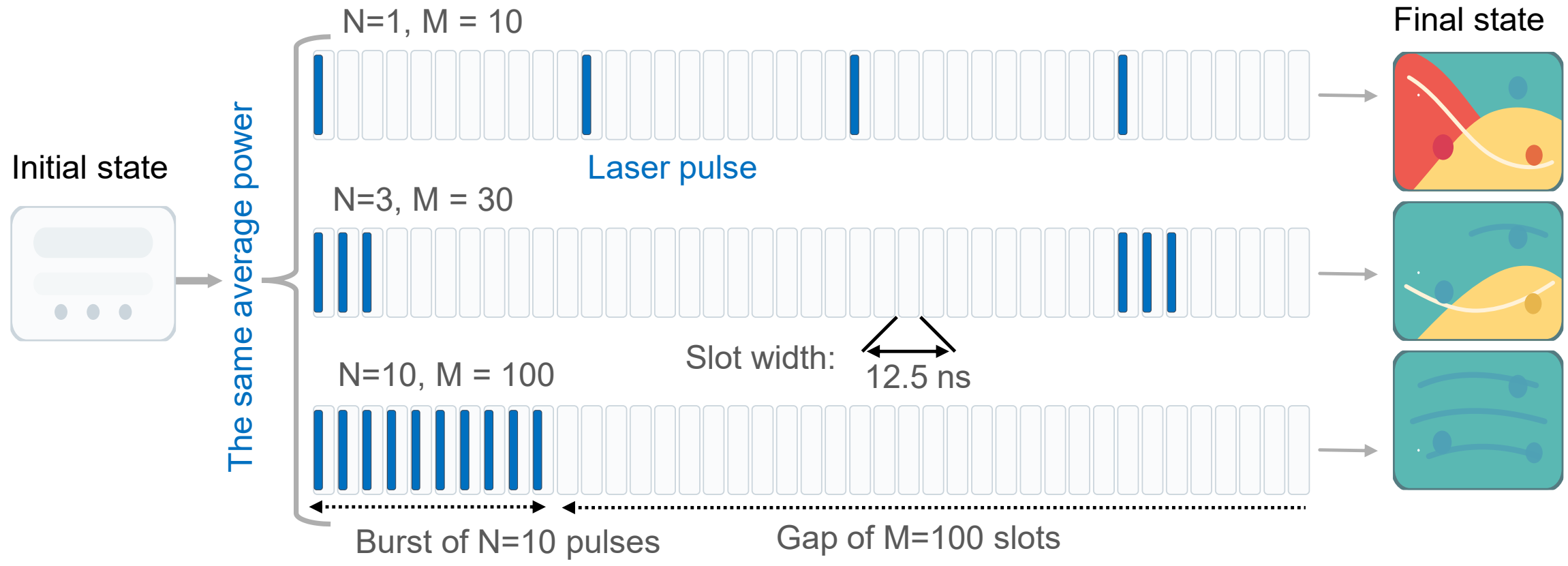

**Figure 1. Explanation of the pulse-burst excitation scheme.** Let us imagine an infinite periodic sequence of time slots, each 12.5 ns long shown as grey boxes in the cartoon. 12.5 ns period corresponds to 80 MHz base frequency. Each time slot either contains a short laser pulse at its start, or not. Setting N consecutive slots with pulses followed by M slots without pulses and repeating this pattern produces a pulse-burst excitation denoted by (N, M). The repetition period of the burst is then (N+M)×12.5 ns. (a) and (b) and (c) show bursts (1,10), (3,30) and (10,100) respectively. Although these burst excitations deliver the same time-average excitation power, they potentially can lead to different terminal photoinduced states of a photosensitive sample as illustrated by the cartoons.

Pulse-burst excitation was generated by triggering a picosecond laser diode.[37] Each of the 12.5 ns time slots (at a base frequency of 80 MHz) could either contain a pulse or remain empty, allowing us to construct periodic laser pulse bursts as illustrated in Figure 1. A pulse-burst excitation comprising N slots with pulses and M empty slots is hereafter denoted by (N, M). For bursts with the same N/M ratio (same duty cycle), changing the burst duration does not impact the average excitation power or photon dose over a given time, as exemplified in Figure 1.

In $MAPbBr_xI_{3-x}$ -type mixed-halide perovskites used here (see Supplementary Note S1), the state of the material probed from the PL spectrum: the mixed state possesses PL centered around 640 nm, while the terminal segregated state emits near 750 nm [7,16,27]. Figure 2a-c shows experimental results obtained on a slightly understoichiometric $MAPbBr(_{0.5}I_{0.5})_{2.96}$ perovskite film[19] under varied burst excitation characterized by the combination (N, M) and a constant single-pulse fluence of 2 $\mu$J cm$^{-2}$ (see Supplementary Note S2).

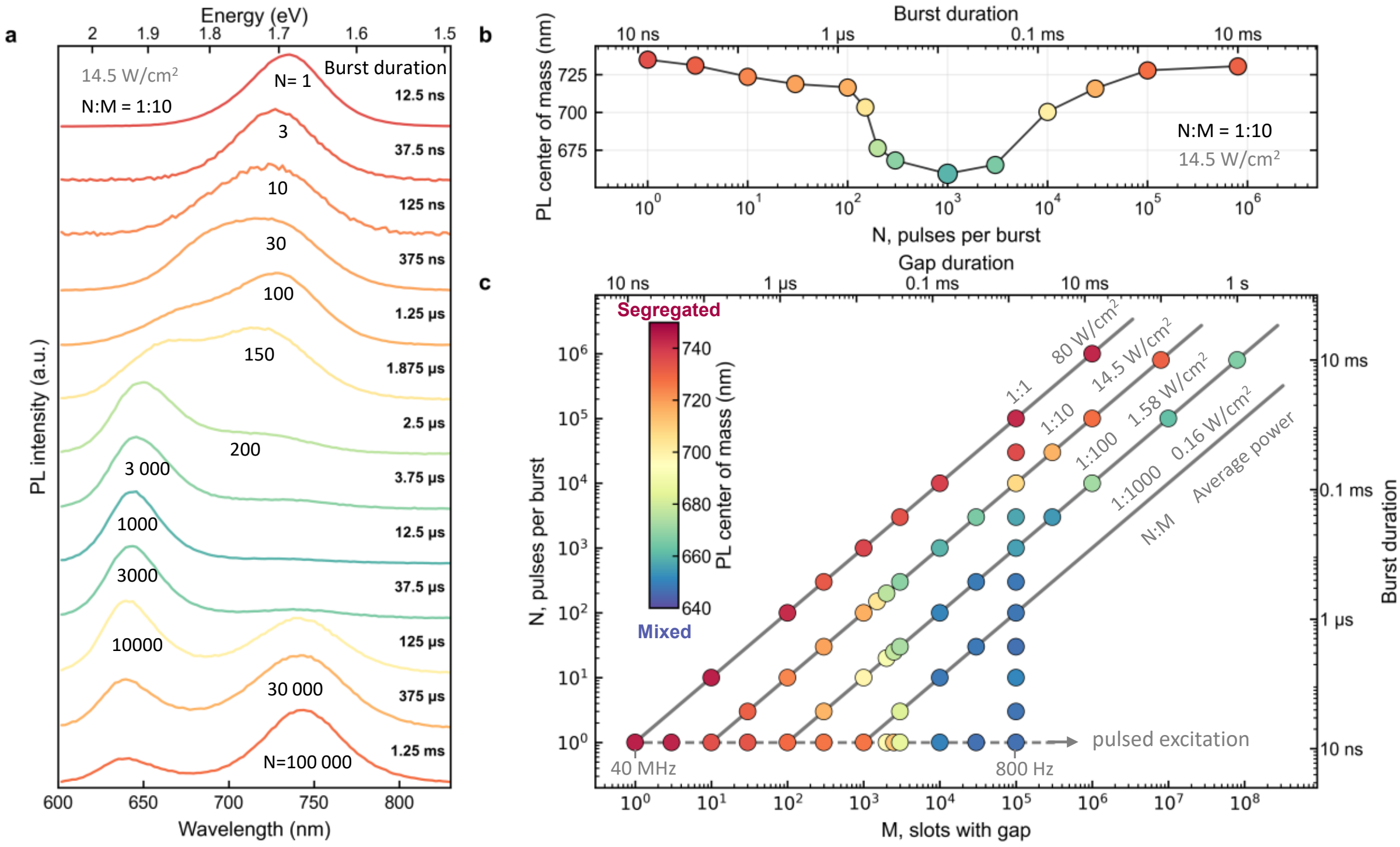

**Figure 2. Controlling of phase segregation state and time-dose reciprocity breakdown in mixed-halide perovskites. (a)** The final (terminal) PL spectra reached after 500 s of exposure to pulse-burst excitation with fixed a duty cycle (M:N=10) and with the number of pulses per burst, N, varied from 1 to 100 000. The average power density and the total dose are constant for all these conditions, while the terminal states are different. **(b)** The PL center of mass extracted from the spectra shown in **(a)** as a function of burst duration (N×12.5 ns). **(c)** Map of the PL center of mass equilibrated after 500 seconds irradiation under different combinations (N,M). Diagonal lines correspond to the excitation conditions with the same photon dose (the same average power density), see the labels. N=1 row corresponds to the conventional pulsed excitation with repetition rates ranging from 40 MHz (M=1) to 800 Hz (M=$10^5$). The colormap is the same for all panels and represents the PL center of mass after 500 seconds for each (N,M) combination.

Figure 2a and 2b show the PL spectra and their centers of mass as a function of N for M:N=10:1, corresponding to the same duty cycle and average power density. Despite the same photon dose, the terminal PL differs strongly. When the bursts are short, the PL peaks near 740 nm, corresponding to an iodide-rich segregated state. Increasing the burst duration shifts the equilibrium toward mixing, and a burst duration of 1.25 µs (N=100) produces an intermediate material state in which PL from the segregated and mixed states is present simultaneously. The burst length in the range 2.5 – 37.5 $\mu s$ clearly brings the sample toward the mixed state. Remarkably, the further increase of the burst length drives the system back to the segregated state. Overall, these data show that the same photon dose applied at different times on a sub-millisecond time scale yields markedly different terminal PL spectra, and that this dependence is strongly nonmonotonic.

To illustrate this, the PL center of mass for these spectra is plotted in Figure 2b, which shows a pronounced U-shaped dependence on the pulse burst duration. Notably, similar center-of-mass values may correspond to different PL spectral shapes, as seen in Figure 2a. Therefore, the center of mass should be viewed as a compact descriptor of the material's emissive state rather than a complete representation of the spectrum. Taken together, both PL spectra and the PL center of mass show that the timing of the pulse burst selects the terminal emissive state reached, given the same photon dose.

The full map (Figure 2c) of PL state in the excitation space for excitation-burst vectors (N,M) confirms that the effect demonstrated above is not an isolated observation. The marked diagonals correspond to conditions with the same photon dose and average power, but the terminal PL center-of-mass shifts from the segregated state toward the mixed state, then back toward the segregated state as N increases. This non-monotonic trend is clearly observed for average power densities of 80 W/cm$^2$ (M:N=1), 14.5 W/cm$^2$ (M:N=10) and 1.58 W/cm$^2$ (M:N=100) (Figure 2c). For 0.16 W/cm$^2$ (M:N=1000), the measured data do not fully capture the return to the segregated state, but such a return is expected at sufficiently long burst durations. This expectation follows from the continuous-wave (CW) excitation limit: at sufficiently long burst lengths, any pulse-burst excitation effectively approaches interrupted CW illumination at 0.16 W/cm$^2$ , which is above the segregation threshold [16,27]. The horizontal N=1 trajectory in Figure 2c corresponds to conventional pulsed excitation with variable repetition rate (different M), and the observed trend in terminal PL spectra agrees with recent observations by Okrepka et al.[17] obtained at a very similar pulse fluence. Overall, the map additionally reveals a general trend: at fixed M, increasing N drives the material toward a more segregated emissive state, whereas at fixed N, increasing M favors a more mixed-like state.

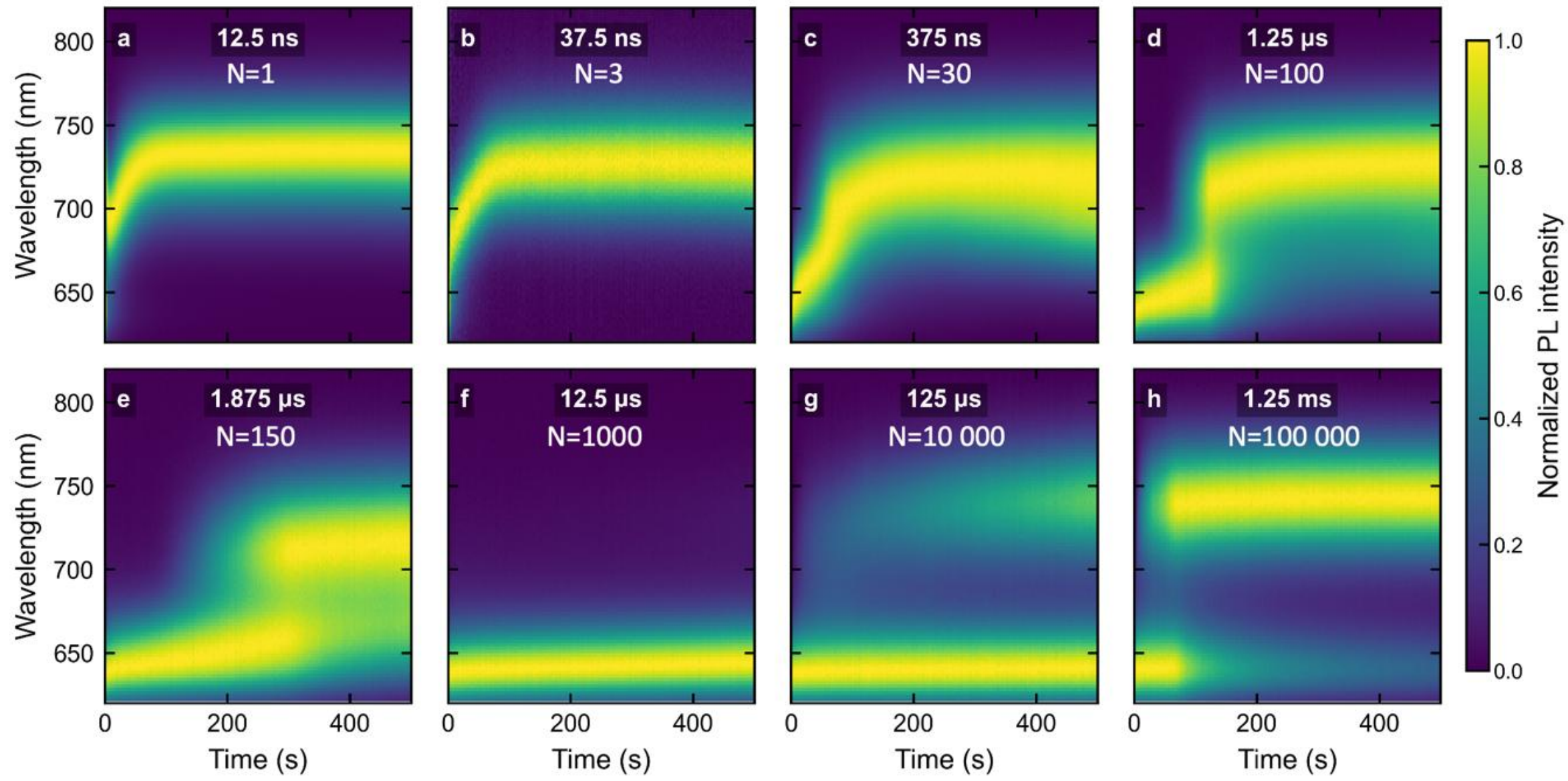

**Figure 3. Time evolution of the normalized photoluminescence spectra under the same average excitation power density but different burst lengths.** N changes from 1 to 100 000 (a-h) while M:N=10 and the average power density is 14.5 W cm$^{-2}$.

After characterizing the terminal PL spectra at different (N,M) conditions (Figure 2), we next ask how these different states are reached. Figure 3 shows the temporal evolution of selected PL spectra for excitation conditions from the (M:N=10) pulse-burst family. Not only do the final spectra differ, but also the dynamics of how these final spectra are reached. Short-duration bursts rapidly drive the sample into a segregation-dominated state (a,b). Longer-duration bursts (375 ns (c), 1.25 $\mu$s (d)) produce a similar effect, but more slowly. However, upon increasing the burst duration tenfold (to 12.5 $\mu$s (f)), the segregation becomes nearly fully suppressed. Upon further increasing the burst length, the 1.25 ms burst rapidly returns the system to the segregated state again on a timescale similar to that for the very short bursts. For this very long burst, however, a visible contribution from the mixed phase remains, and the mixed phase appears in a step-like manner, in contrast to the gradual shift observed for the short bursts. The spectral dynamics for all (N, M) pulse-burst combinations shown in the map in Figure 2c are given in the SI in Figures S1 and S2.

Let us look closely at the regime in which the segregation dynamics are especially sensitive to the burst length. A relatively small change in burst duration from 1.875 $\mu s$ (N=150) to 12.5 $\mu s$ (N=1000) leads to completely different final PL states (Figure 3). Note that all these timescales include the corresponding burst gaps (from 18.75 $\mu s$ to 125 $\mu s$), which are much shorter than the characteristic timescales of segregation (tens of seconds). At the same time, these gaps are much longer than the PL lifetime in our samples – approximately 10 ns at the given pulse fluence – which reflects the decay of at least one type of free charge carrier, electron or hole (Figure S3). If we consider a simple picture of ion segregation driven by charge carrier dynamics[17,27] when the charge carrier lifetime is negligible compared with the dark gap, excitation patterns with gaps much longer than the PL lifetime should behave similarly. The fact that they do not, as shown in Figure 3, indicates that additional slower processes must be involved, see the discussion section below.

We further ask how controllable these photochemical transformations are. To test this, we probed the reversibility of switching between two burst conditions with the same duty cycle (the same M:N). Based on the (N, M) map in Figure 2c, we chose the (100, 10000) burst excitation to promote ion mixing, while the (1, 100) burst condition was used to promote ion segregation. The evolution of the PL spectra under these excitation patterns is shown in Figure 4. While the first several tens of cycles appear fully reversible, slower changes in the material led to a different quasi-steady-state pattern after approximately 100 cycles (Figure 4b). The switching in the last three cycles (after ≈1000 cycles run) is clearly different from that of the first three cycles, yet reversible PL modulation remains evident even at the end of this 20-hour-long experiment. Interestingly, the terminal material state reached after 20 hours of illumination (Figure 4c) appears to be much more stable in switching repeatability and dark stability than the initial state of the film before illumination. We checked this by allowing the sample to rest in the dark for 60 min and testing switching cycles again (Figure 4d). The switching patterns before and after resting are very similar (compare panels c and d) and do not show any drift dynamics typical for the initial stage of the experiment (compare panels a and b). Thus, temporal structuring of excitation provides a means to reversibly switch the material between distinct photoinduced states without changing the total photon dose. The slow, irreversible transformations of the material might arise from chemical reactions such as iodide oxidation, followed by the loss of gaseous $I_2$ [43,44]. Nevertheless, even after this long-term exposure, the material could still be reversibly manipulated by switching between the two pulse-burst conditions.

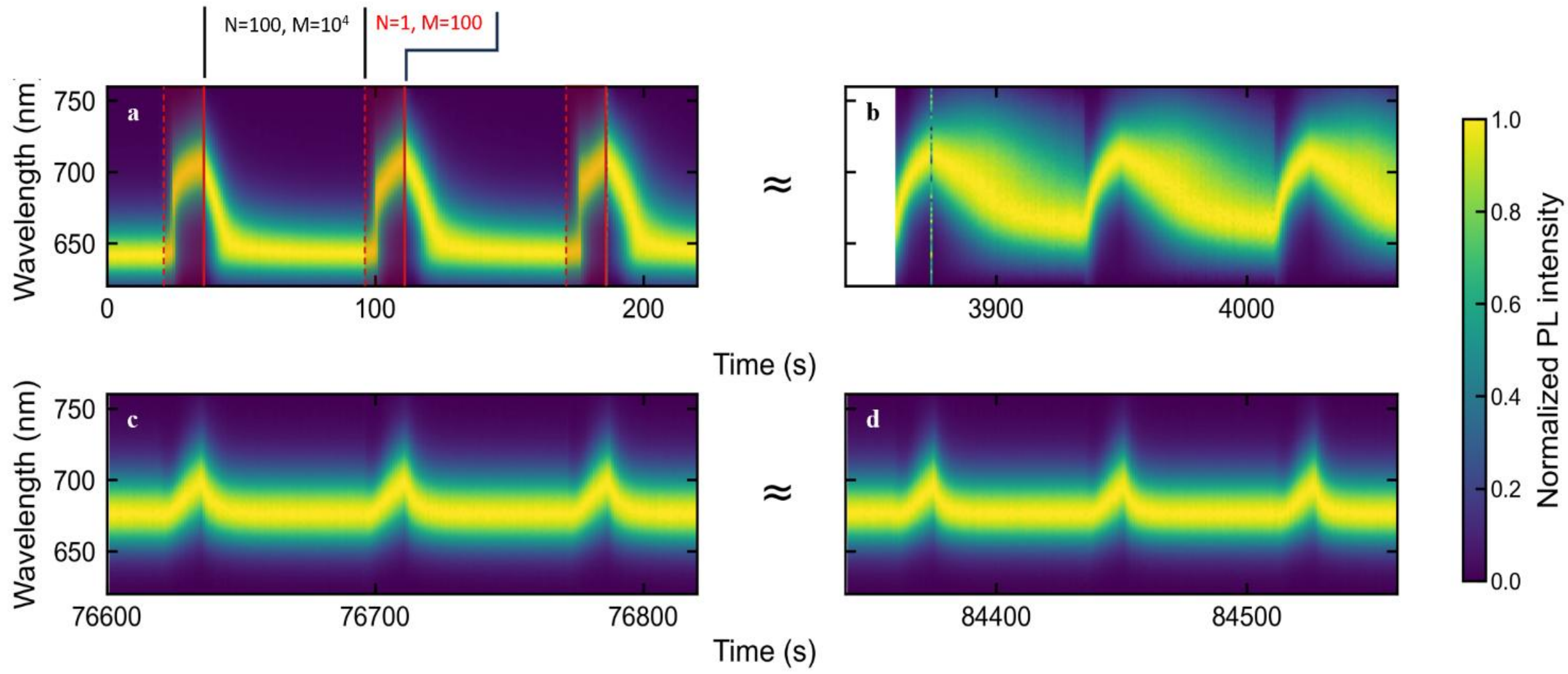

**Figure 4. Reversible switching between segregation and mixing over more than 1000 cycles.** The switching performed using a segregation-driving burst condition (N=1, M =100) for 15 s, and a mixing-driving burst condition (N=100, M=$10^4$) for 60 s. **(a)** First three cycles. **(b)** Three cycles after 1 h of measurement. **(c)** Last three cycles. **(d)** Switching after 60 min in the dark. The average power density is 1.58 W $cm^{-2}$. PL is normalized in each spectrum.

**Discussion**

Trapped charges in various metal-halide perovskites have been shown to persist for timescales ranging from microseconds to tens or even hundreds of microseconds [37–39]. If trapping is asymmetric, with one type of charge carrier preferentially captured, for example, electrons (n), this leads to photodoping. As a result, free charge carriers (p) of the opposite type can remain in the material for hundreds of microseconds (e.g., holes, with $p \gg n$), together with trapped charges of the other type (electrons in this example). Although this effect has been widely reported and analyzed for several classical metal-halide perovskites, it has rarely, if ever, been considered in theoretical models of photoinduced halide segregation in mixed-halide systems, where the densities of free electrons and holes are typically not treated separately, thus assumed to be equal (n=p).

The depopulation of trapped charges, and therefore the decay of photodoping, is essentially non-exponential with a long tail. In the simplest model, it follows a hyperbolic decay:[37]

$$n_t = \frac{{n_t}^0}{(1+k_n {n_t}^0\, t)} \qquad (1)$$

where $k_n$ is the trap depopulation rate constant, ${n_t}^0$ is the concentration of trapped charges immediately after excitation (t=0), and $n_t$ is the concentration of trapped charges at a time t after the end of excitation. The characteristic time of this decay is $1/k_n {n_t}^0$ when the concentration of trapped charges decreases by a factor of two. Importantly, this characteristic time increases as the decay proceeds, since it is inversely proportional to the concentration.

The detection of preferential trapping of one charge-carrier type can be achieved by observing PL properties modulated by the photodoping effect.[45,29,40,37] Specially designed for this purpose, time-resolved PL measurements under multi-pulse excitation in the so-called read-write-read scheme proposed by Marunchenko et al [37] were carried out for our samples in the fully mixed states and fully segregated states as described in Supplementary Note S5. Applying the equation above to the analysis, $k_n$ was estimated to be $7.7\times10^{-10}$ and $1.8\times10^{-10}$ $cm^3s^{-1}$ for the fully mixed and segregated states, respectively, see Supplementary Note S5 for details. These values define an approximate $k_n$ range within which intermediate material states are expected to lie.

Knowing $k_n$ allows estimation of the population of trapped charge carriers at any time after the burst excitation. In our experiments, each pulse generates $10^{17}$ $cm^{-3}$ photoexcited carriers. In the case of a strong photodoping effect, most of these charge carriers are trapped shortly after the excitation. Continuous excitation by N pulses in the burst can potentially create an even larger trapped-carrier population if enough traps are available. According to Eq. 1, taking $k_n$ $=2\times10^{-10}$ cm$^3$s$^{-1}$, an initial trapped-charge concentration of $10^{17}$ cm$^{-3}$ at the end of the burst (N > 1) decreases to approximately $10^{13}$ cm$^{-3}$ after 1 ms (this time corresponds to M≈$10^5$). This residual concentration is substantial and comparable to the theoretically estimated carrier-density threshold for light-induced phase segregation in $MAPbBr(_{0.5}I_{0.5})_3$, which is on the order of $10^{13}$ – $10^{14}$ cm$^{-3}$.[27] Consequently, the temporal structure of the excitation is particularly important on timescales relevant to trapped-charge relaxation. Persistent trapped charges can sustain photodoping between successive bursts, thereby contributing to phase segregation.

To summarize, each pulse burst creates long-lived hidden photoexcitations in the material. One such perturbation is the population of long-lived trapped charges and the associated photodoping discussed here; others may include the buildup of local electric fields[12], strain [41,42], and temperature.[11] Each subsequent dark interval allows these hidden states to relax, thereby modifying the initial conditions for the next burst and ultimately influencing the metastable compositional state reached by the material. The observed PL evolution is therefore the cumulative result of repeated excitation–relaxation cycles involving these hidden variables, rather than a direct response to photon dose alone.

The observed dynamics in mixed-halide semiconductors resemble classic intermittent-light experiments in photosensitive systems. For example, in Schwarzschild's intermittent-exposure experiment, splitting the same nominal light dose into different temporal intervals leads to a different degree of photodarkening in the image obtained on silver bromide gelatine emulsion[3]. In photosynthesis, flash-light experiments were performed to separate fast photochemical excitation from slower dark reactions, revealing that the yield depends on the timing between flashes[46]. In the case of mixed-halide perovskites, the pulse burst acts as the flash, the dark gap as the recovery interval, and the terminal PL spectrum as a probe of photoactivated intermediates that can accumulate, saturate, or relax.

To the best of our knowledge, the only published work in which an interrupted excitation was used for mixed halides is the report by Knight et al.[12], which investigated $MAPb(Br_{0.5}I_{0.5})_3$ under CW laser irradiation. They specifically found that periodic interruption of the CW laser (0.06 W/cm$^2$) at frequencies ranging from 10 to 3500 Hz while keeping the fixed duty cycle at 50% (N=M in our terminology) did not change the segregation dynamics. It was concluded that modulation frequency alone does not directly affect halide segregation. In the current work, we show that this conclusion holds only for the selected excitation conditions, whereas a dependence on modulation frequency emerges when a broader range in N, M, and average excitation power density is probed.

In our experiments, the dark intervals allow the sample to relax between pulse bursts. A particularly interesting and counterintuitive phenomenon observed in mixed-halide perovskites is light-induced mixing or photoremixing,[14,15,28] in which the transition from the segregated to the mixed state can proceed faster under certain excitation conditions than in the dark. A natural extension of the pulse-burst experiment would therefore be to replace the dark intervals with a tailored low-repetition-rate pulsed excitation that actively drives the material toward the mixed state. This would introduce an additional control parameter for tuning the competition between segregation and remixing, and hence the terminal state of the material. Such generalized pulse burst experiments could help existing theoretical approaches toward developing a unified description of photoinduced segregation and remixing in mixed-halide perovskites [27,21,47].

The practical significance of our results is that pulse bursts can be used to manipulate perovskite photochemistry over broad timescales with several potential applications. For example, temporal structuring of the excitation can substantially expand the illumination conditions under which mixed-halide perovskites remain in the mixed state without phase segregation. According to Figure 2c, at a single-pulse fluence of 2 μJ/cm$^2$, the material remains in the mixed state for burst widths of approximately 1.25–12.5 μs ($N = 100$–$1000$), despite the rather strong excitation. Moreover, under N=100, M=$10^4$ excitation, the PL spectrum and intensity remain nearly unchanged for 24 h (Figure S7). This reveals previously unexplored operating windows in the excitation landscape of mixed-halide perovskites, in which temporally structured excitation can stabilize the mixed state under conditions where segregation might otherwise be expected.

Temporal structuring of excitation can serve as input to program the PL spectrum of mixed halides. Very recently, Ruth et al.[28] demonstrated spectral tuning of mixed-halide perovskites by varying pulsed excitation parameters such as repetition rate and peak fluence. Kouwenhoven et al. [48] further exploited excitation-power- and repetition-rate-dependent PL wavelength shifts to realize a reconfigurable multistate optical memory, in which the PL spectrum encodes the history of optical excitation, and the continuously tunable degree of segregation can represent an analog synaptic weight. Our switching experiment (Figure 4) isolates a different control coordinate – the burst length. The N=1, M=100 condition corresponds to isolated pulses at an effective repetition rate of 792 kHz, whereas N=100, M=$10^4$ delivers groups of 100 closely spaced pulses at 80 MHz, separated by much longer dark intervals. Both conditions have essentially the same duty cycle, single-pulse fluence, average power, and photon dose, yet they produce distinctly different emission colors. Thus, the temporal packaging of otherwise identical photons provides an additional degree of freedom for programming the material state. This excitation-history-dependent response places mixed-halide perovskites within the broader memlumor concept [49], in which the PL of a material depends on an internal state that carries a memory of previous excitation. Here, the halide composition itself serves as a slowly evolving memory variable that can be addressed through the temporal structure of the incident light.

**Conclusion.**

To conclude, we show that temporally structured nonequilibrium photoexcitation provides an independent control parameter for selecting the state of mixed-halide perovskites. Varying the pulse-burst timing while keeping the single-pulse fluence, average power, and photon dose fixed drives the material into distinctly different metastable emissive states. The pronounced nonmonotonic dependence of the final state on burst duration demonstrates that neither photon dose nor average power alone determines the material response. Our measurements further reveal long-lived trapped-charge populations and associated photodoping as one mechanism capable of retaining memory between consecutive bursts. By modifying the electronic state from which each subsequent burst acts, such or similar long-lived populations can influence the balance between halide segregation and mixing. Mixed-halide perovskites therefore retain information about photon arrival times, allowing their PL emission spectra and intensity to be controlled through photon timing and making this excitation-history-dependent response particularly relevant to neuromorphic photonic applications.

**Author contributions**

A.M. and I.S. conceived the project idea. A.M. designed the experiments and performed the measurements. D.L. developed the custom automated program for pulse-burst control. B.D. developed temporal separation of the CW laser excitation and the TCSPC detection, and did the RWR measurements. S.S. fabricated the samples. A.M. processed all data. A.M. and I.S.

wrote the manuscript with contributions from all authors. I.S. and Y.V. supervised the project. All authors contributed to the discussions and commented on the paper.

**Acknowledgements**
A.M. acknowledges the Light and Materials profile area at Lund University (Young Investigator Synergy Award). The study was supported by NanoLund (12-2023) and the Crafoord Foundation (20230552). B.D. thanks the Carl Trygger Foundation for a postdoctoral scholarship. Funding by the German Research Foundation (Deutsche Forschungsgemeinschaft, DFG) *via* the "Responsible Electronics in the Climate Change Era – REC$^2$" Cluster of Excellence (EXC 3035, Project-ID 533607596) is gratefully acknowledged.

# Supplementary Information for: Pulse-Burst Excitation Reveals Time–Dose Reciprocity Breakdown in Mixed-Halide Perovskites

Alexandr Marunchenko[1], Shivam Singh [2,3], Daniel Lizotte[1], Bhaskar De[1], Yana Vaynzof [2,3], Ivan G. Scheblykin[1*]

[1] Chemical Physics and NanoLund, Lund University, P.O. Box 124, 22100 Lund, Sweden

[2] Chair for Emerging Electronic Technologies, Technical University of Dresden, Nöthnitzer Str. 61, 01187 Dresden, Germany

[3] Leibniz Institute for Solid State and Materials Research Dresden, Helmholtzstraße 20, 01069 Dresden, Germany

## Supplementary Note S1: Sample preparation

**Materials.** Lead acetate trihydrate [$Pb(Ac)_2 \cdot 3H_2O$], dimethylformamide (DMF) and hypophosphorous acid (HPA) were purchased from Sigma-Aldrich. Methylammonium iodide (MAI) and methylammonium bromide (MABr) were purchased from TCI chemicals. All the materials were used as received.
**Solution preparation.** 40 wt. % $CH_3NH_3PbI_3$ and $CH_3NH_3PbBr_3$ perovskite precursor solutions were prepared using $Pb(Ac)_2 \cdot 3H_2O$ and MAI or MABr in the molar ratio of 1:2.96 in dimethylformamide (DMF), respectively. 6 $\mu$l HPA was added to 1 ml of both the pure halide perovskite precursor solutions, separately, and stirred at room temperature for 4 h. The two pure halide-based perovskite solutions were mixed in the volumetric ratio of 1:1 to form $CH_3NH_3Pb(I_{1.5}Br_{1.5})_{2.96}$ perovskite precursor solutions. The mixed-halide-based perovskite solutions were stirred for another 1 h at room temperature before use.
**Perovskite film fabrication.** The perovskite thin films were fabricated on quartz substrates (purchased from Thorlabs). The substrates were cleaned sequentially with soap solution, deionized water, acetone, and isopropanol for 10 min in an ultrasonication bath. The substrates were then dried with $N_2$ and treated with oxygen plasma at 100 mW for 10 min. The substrates were immediately transferred to a drybox (relative humidity < 1%) for perovskite deposition. The mixed halide perovskite precursor solutions were spin-coated on the quartz substrate at 4000 rpm for 30 s with an acceleration of 1000 rpm/s. The spin-coated perovskite films were dried at room temperature for 5 min on the workbench and then annealed at 100 °C on a hot plate for 5 min, yielding films with a thickness of ~270 nm. To protect against environmental degradation, the perovskite films were coated using PMMA. PMMA was dissolved at a concentration of 10mg/mL in chlorobenzene and cast at 4000 rpm for 30 seconds.

## Supplementary Note S2: Optical setup

Photoluminescence (PL) was measured using a home-built wide-field photoluminescence microscope system based on Olympus IX71. A 485 nm diode laser (PicoQuant, with a 200 ps pulse width) through a 40x objective lens was used[S1]. The size of the excitation spot was around 30 µm x 30 µm. The pulse fluence was set to 2 $\mu$J/cm$^2$ for all measurements from main text, providing the estimated initial carrier concentration of photogenerated electrons and holes $n_0$ =

$p_0 \sim 10^{17}$ cm$^{-3}$. Fluence could be reduced by factor of 10 or 100 using additional optical density filters. For generating pulse bursts, the laser was controlled by the Sepia Oscillator Module (SOM 828-D, PicoQuant). For more details on control of the picosecond laser diode with the SOM 828-D system, see the Supplementary materials in the reference[S1]. Switching between different burst generations during long experiments through SOM 828-D has been realized via a custom software. The spectra were obtained using a slit and a transmission diffraction grating with 150 grooves per mm. For making time-resolved decay measurements, the hybrid photomultiplier detector (PMA Hybrid-42 PicoQuant) was connected to a time-correlated single photon counting (TCSPC) module (PicoHarp 300 PicoQuant) for measurements of PL decay kinetics. The instrumental response function of the TCSPC was approximately 200 ps.

To reduce the dependence of the results on the history of previous measurements, each new measurement was performed at an unexposed location on the film. In our experiments, absolute PL intensities can vary because of local film inhomogeneities. Therefore, we focus on spectral PL observables, such as PL shape and spectral center mass, rather than PL intensity alone.

### Supplementary Note S3: Spectral evolution dynamics for all N, M combinations

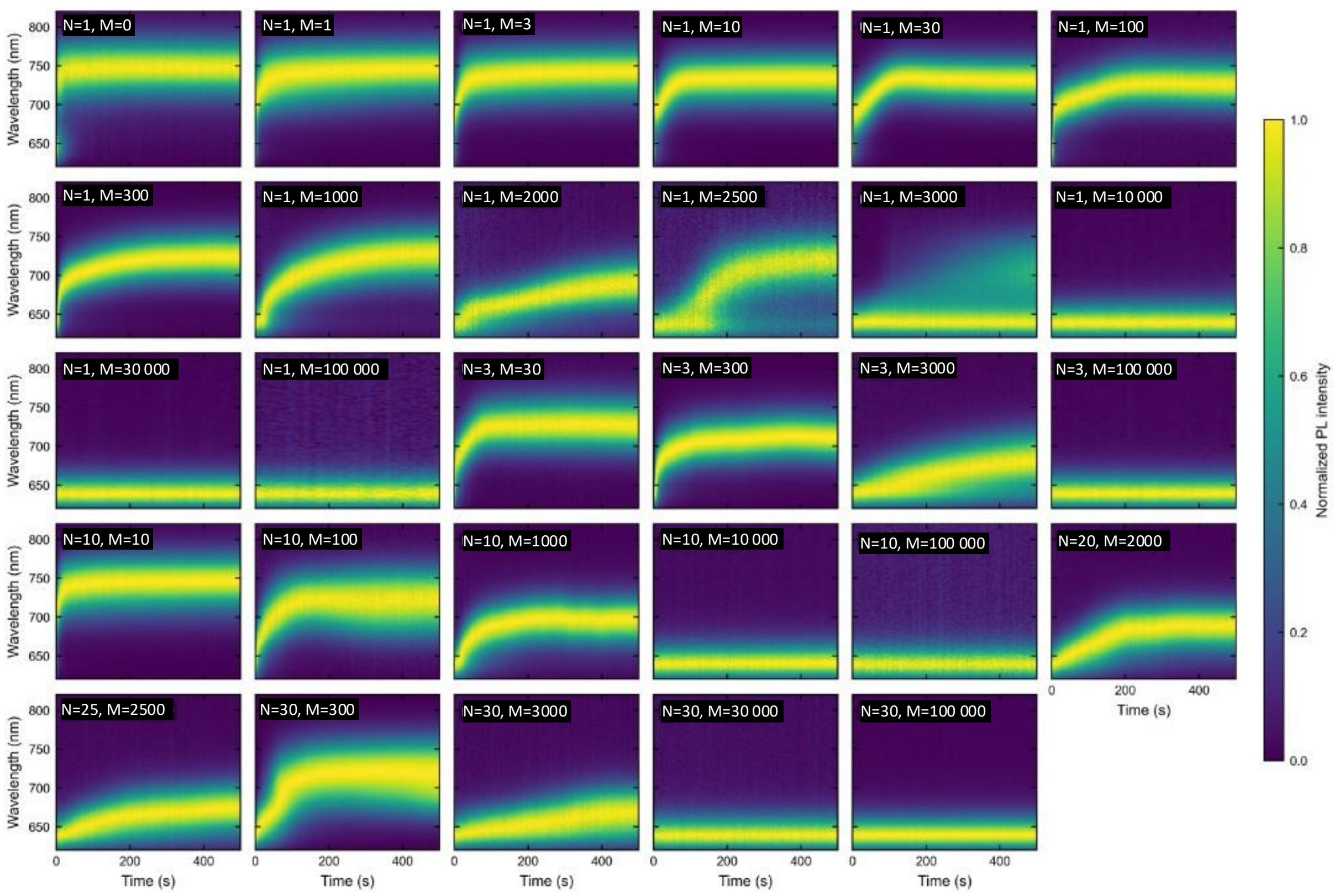

**Figure S1.** PL spectral evolution under pulse burst excitation for combinations of N and M with N<100. The combination N=1, M=0 corresponds to a pulse repetition rate of 80 MHz, whereas N =1, M=1 corresponds to a pulse repetition rate of 40 MHz. Intensity was normalized in each spectrum.

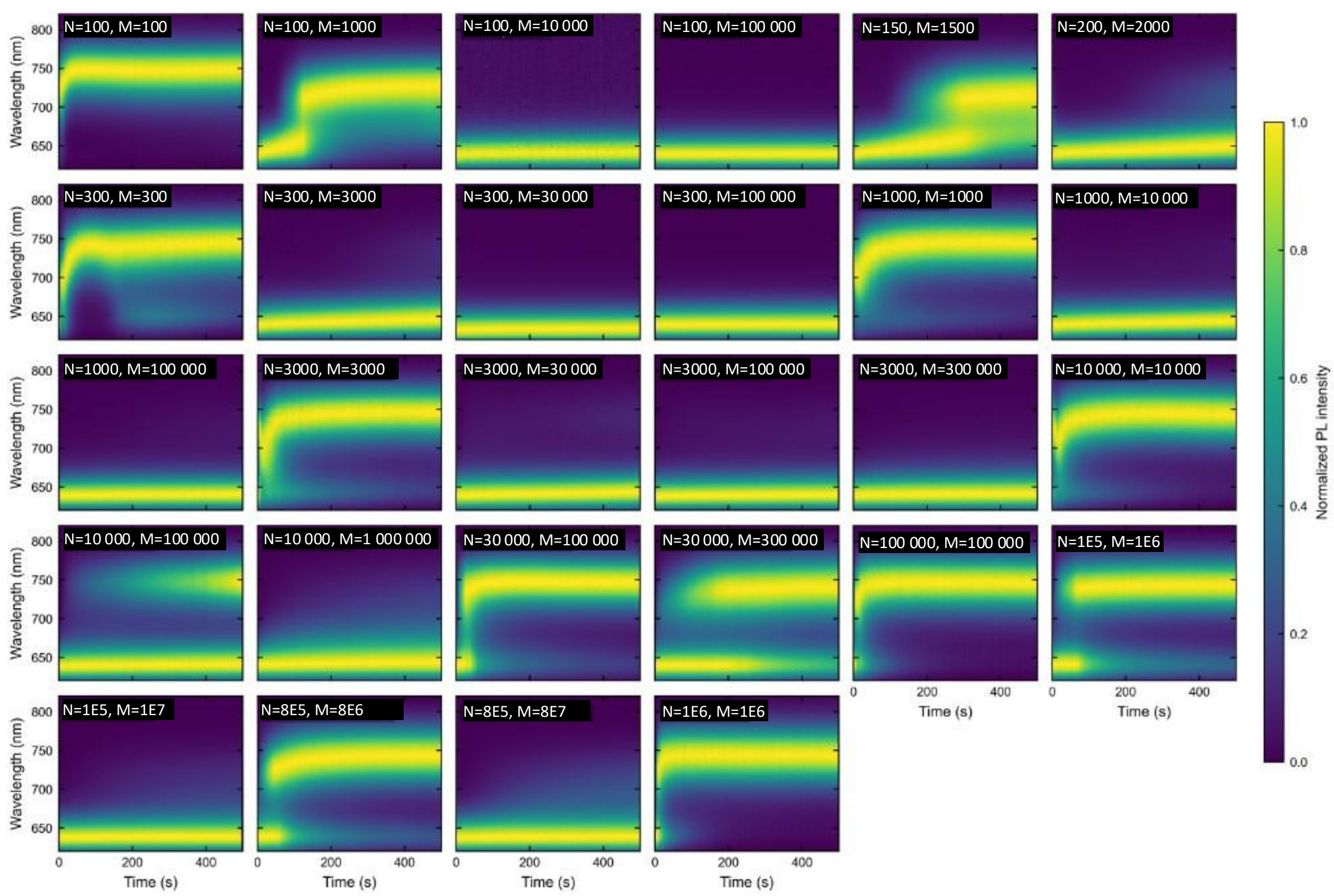

**Figure S2.** PL spectral evolution under pulse burst excitation for combinations of N and M when N≥100. Intensity was normalized in each spectrum.

## Supplementary Note S4: TRPL under selected pulse burst excitation and high fluence

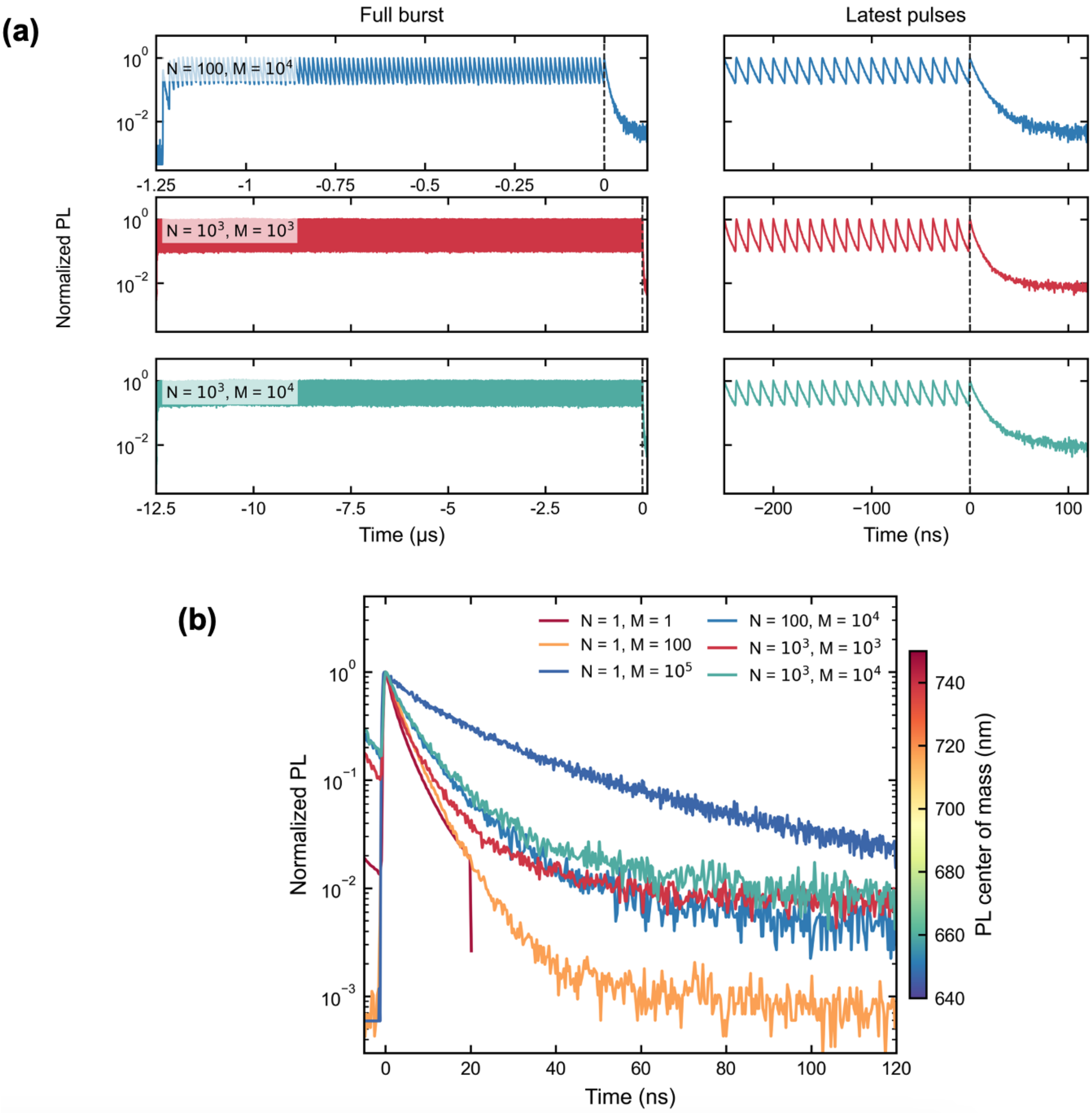

**Figure S3. (a)** Time-resolved photoluminescence (TRPL) response to burst excitations with different N and M combinations. Left: full PL dynamics; right: PL response to the last several pulses of each burst. (b) PL decay after the excitation by the last pulse of the burst for very different N, M combinations. Among the measured combinations there are conditions correspond to conventional pulsed excitation at repetition rates of 80 MHz (1,0), 792 kHz (1,100) and 800 Hz (1,100000). To highlight differences in material state, the line colors indicate the PL spectral center of mass according to the color bar: red represents the segregated state, whereas dark blue represents the mixed state. The effective lifetime (1/e drop of intensity) for most of the decays is about 10 ns or less. For 800 Hz repetition rate (N=1, M=100 000), the lifetime is slightly longer. All measurements were performed at a fluence of 2 $\mu$J cm$^{-2}$.

**Supplementary Note S5: Read-Write-Read TRPL experiments in the mixed and segregated state and estimation of the recombination constant of trapped charges.**

For PL excitation in the Read-Write-Read scheme (RWR)[S1], we used the same laser (485 nm) as for controlling segregation, but with a pulse fluence approximately 100 times lower (~0.02 μJ/cm$^2$). The repetition period of the sequence was 25 μs. We performed the RWR experiment by applying a first read pulse ($R_1$), followed by a write burst (W) and a second read pulse ($R_2$). The delay between $R_1$ and W was fixed at 0.5 μs, while the delay ($t_3$) between W and $R_2$ was varied. The write burst consisted of 20 pulses @ 40 MHz (25 ns between pulses).

Under this condition, initially, the mixed state did not change during the data acquisition (typically 200 s). The status of the sample was monitored by observing the emission at ~640 nm. The TRPL results are shown in Figure S4.

The segregated state, however, was unstable under this burst excitation and quickly shifted toward the mixed state on a timescale of seconds. To stabilize the segregated state under strong irradiation and, at the same time, to measure RWR at low irradiation, we used interrupted stabilization, as described below.

The segregated state was prepared by illuminating the sample with a CW Ar-ion laser (488 nm) with a power density of ~140 W/cm$^2$. The sample was continuously illuminated with the pulsed laser (same parameters as described above), while illumination with the CW laser was interrupted using an electrical shutter ($shutter_1$). Another shutter ($shutter_2$) was installed in front of the TRPL detection system. Using LabVIEW software, we could sequentially open/close both shutters so that when $shutter_1$ was open (CW laser was hitting the sample), $shutter_2$ (TRPL detection) was closed, and *vice versa*. This blocked photon counting during CW laser irradiation. The open time for $shutter_1$ (CW laser) was 0.5 s, while for $shutter_2$ (detection) it was 1 s. Excitation by the pulse burst sequence (the same as for the mixed state described above) was constantly applied to the sample. The results are shown in Figure S5.

Panels (c) in both figures show the analysis of the data to estimate the rate constant of trap state depopulation ($k_n$) according to the method from Ref. S1. This method and the theory behind it will shortly be described below based on the full derivation presented previously.

Decay of initially created trap carrier population in the model with one defect level follows the following hyperbolic functions:

$$n_t = \frac{n_t{}^0}{(1+k_n n_t{}^0\, t\,)} \qquad \text{(S1)}$$

where $k_n$ is the trap depopulation rate constant, $n_t{}^0$ is the concentration of trapped charges immediately after excitation (t=0), and $n_t$ is the concentration of trapped charges, and t is the time after the end of excitation. See SI of Reference S1, for understanding limitations, and conditions at which this equation was derived.

In the RWR experiment, the first $R_1$ pulse probes the sample in its ground state in terms of the presence of photodoping: $n_t$(before $R_1$ arrival)=0. Therefore, the PL response to this initial pulse (initial PL amplitude) is proportional to the square of the initial photogenerated charge carrier concentration: $n_o^2$. After that, the response to each subsequent pulse will be influenced by the photodoping generated by previous pulses. Assuming strong photodoping due to e.g. trapping of electrons (decay of electron population is very fast, decay of $n_t$ and p is very slow),

PL intensity is proportional to $np \approx n_0 n_t$ because $n+n_t=p$ and $n \ll p$. Thus, the initial PL amplitude of the response to the $R_2$ pulse is proportional to the residual $n_t$ present in the sample just before the $R_2$ pulse arrived. The reference in this experiment is the PL response to the $R_1$ pulse, which reflects the signal without photodoping yet. Note that here we always refer to the initial amplitude of the time-resolved PL response to a pulse. Taking this together, according to Ref. S1:

$$n_t = n_0 \frac{PL(R_2) - PL(R_1)}{PL(R_1)} = n_0 \frac{\Delta PL}{PL(R_1)} \tag{S2}$$

We did have strong photodoping in our samples, as Figure S5 shows a substantial increase in the response to each subsequent pulse within the pulse burst (by response, we mean here the increase in PL due to the arrival of an excitation pulse) relative to the initial response to pulse R1. This growth is about 5-8 times for both mixed and segregated states of the sample. This means that the photodoping decay (or decay of trapped carriers) is much slower than the distance between pulses. Moreover, the residual PL just before the arrival of the next pulse is rather small in comparison with the PL initiated by the pulse, meaning that PL decay is shorter than the distance of 25 ns between pulses. In these properties, our mixed-halide samples were rather close to the $CsPbBr_3$ microwires discussed in Ref.S1.

Using Eq. S2 and assuming $n_0=1.9\times10^{15}$ cm$^{-3}$ (0.02 μJ/cm$^2$ pulse fully absorbed and distributed over 270 nm thickness of the film), the RWR experiments were analyzed as shown in Figure S4c and S5c. The fitting gave $k_n\approx8\times10^{-10}$ cm$^3$s$^{-1}$ for the mixed states, $k_n\approx2\times10^{-10}$ cm$^3$s$^{-1}$ for the segregated state of the sample.

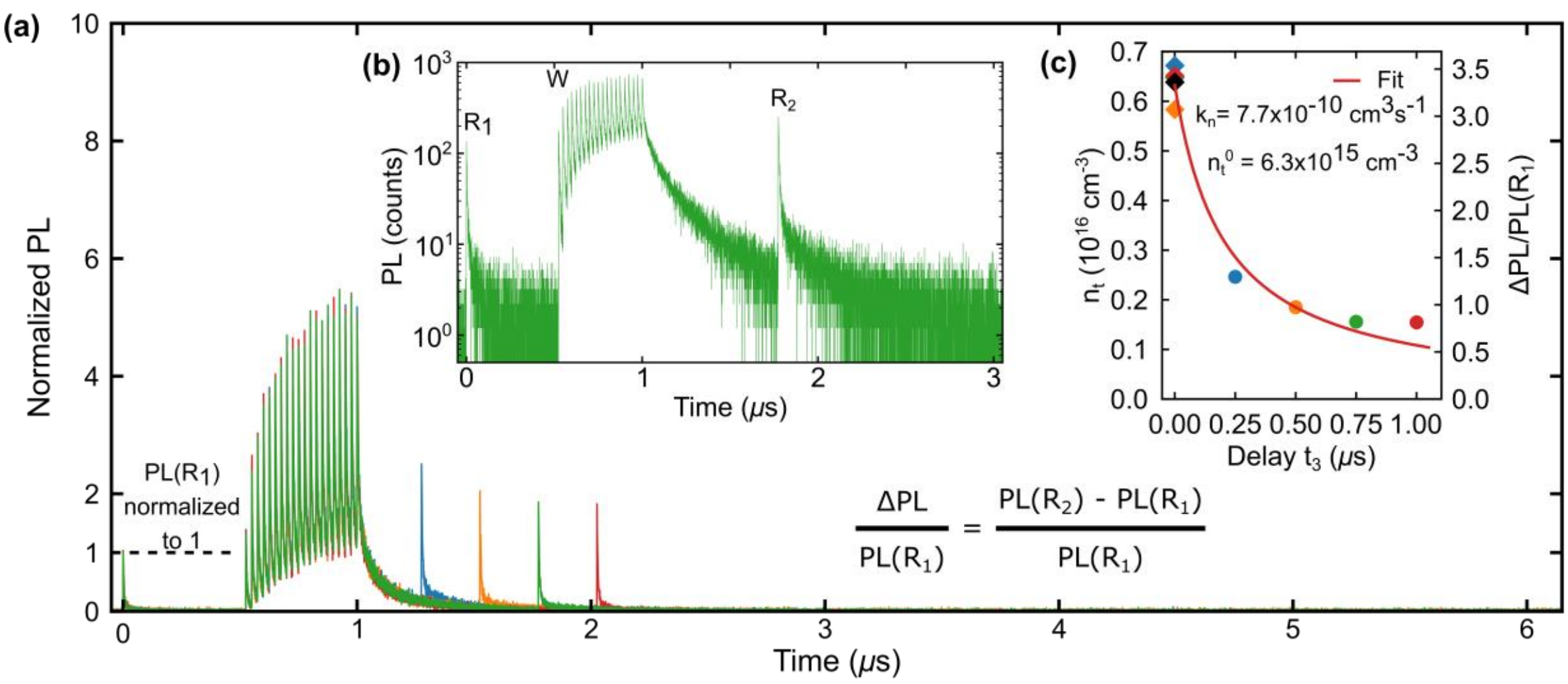

**Figure S4.** (a) TRPL response of the mixed state of the sample to the RWR excitation pulse sequence ($R_1$ pulse, gap, 20 pulse burst (W), $R_2$ pulse) for different delay $t_3$ of $R_2$ pulse from the end of the burst (shown by different colors). Each curve is normalized to the response to the first read pulse, PL($R_1$) =1. The delays between $R_1$ and W were the same, as well as the burst length (W) thus the PL responses are overlapping in the figure except for the responses to $R_2$ pulses. (b) TRPL response for $t_3$=0.75 μs shown on a log scale. (c) The decay of PL memory $\Delta PL/PL(R_1)$ as a function of delay $t_3$. Before calculating the PL amplitudes, a background noise and background luminescence that had not yet fully decayed from the pulse burst were subtracted. The latter was done using a spline fitted to the tail. The colors of the data points are given in the same color code as used for the TRPL signals in (a). Square symbols show the data for the last pulse in the burst, which can be seen as a read pulse for $t_3$=0; the black square is the average value.

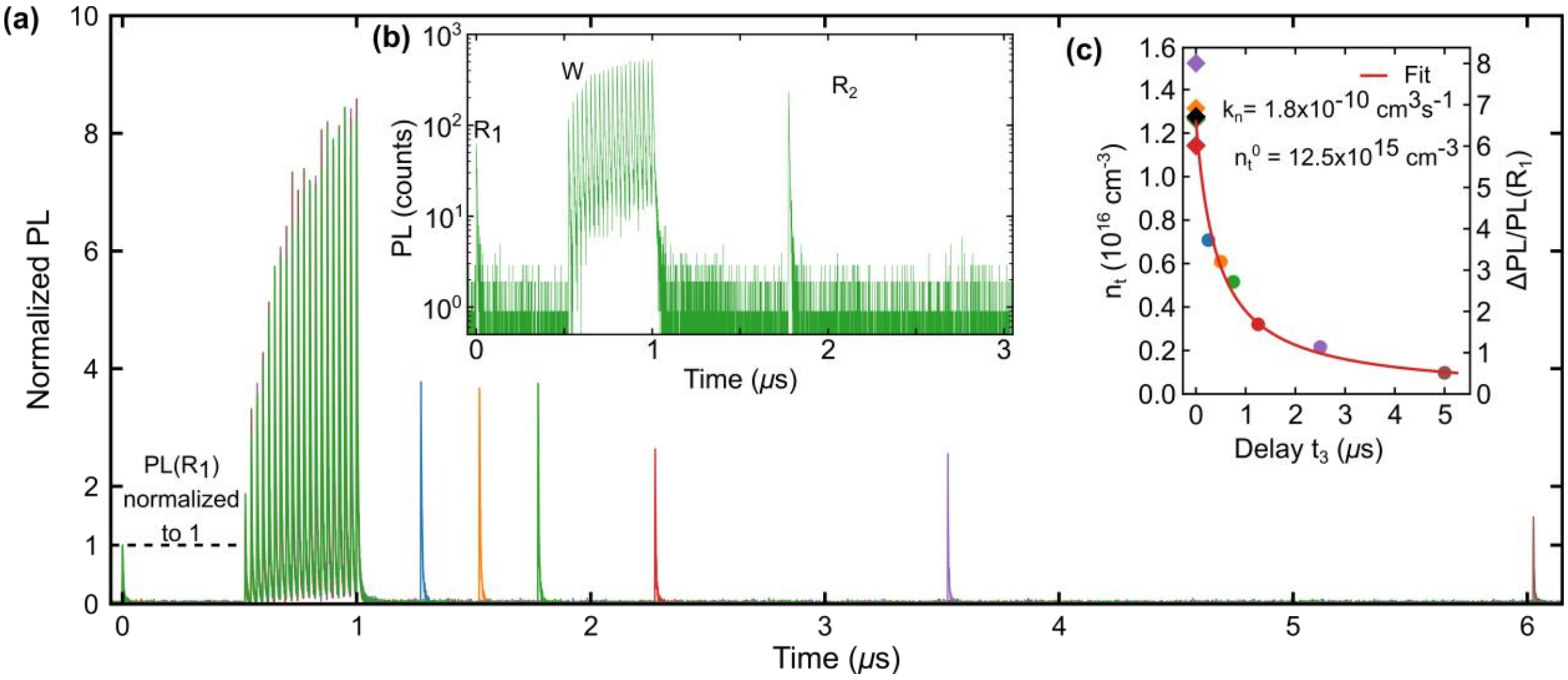

**Figure S5**. (a) TRPL response of the segregated state of the sample to the RWR excitation pulse sequence ($R_1$ pulse, gap, 20 pulse burst (W), $R_2$ pulse) for different delay $t_3$ of $R_2$ pulse from the end of the burst (shown by different colors). Each curve is normalized to the response to the first read pulse, PL($R_1$) =1. (b) TRPL response for $t_3$=0.75 μs shown on a log scale. (c) The decay of PL memory $\Delta PL/PL(R_1)$ as a function of delay $t_3$. Before calculating the PL amplitudes, a background noise was subtracted. Square symbols show the data for the last pulse in the burst, which can be seen as a read pulse for $t_3$=0; the black square is the average value.

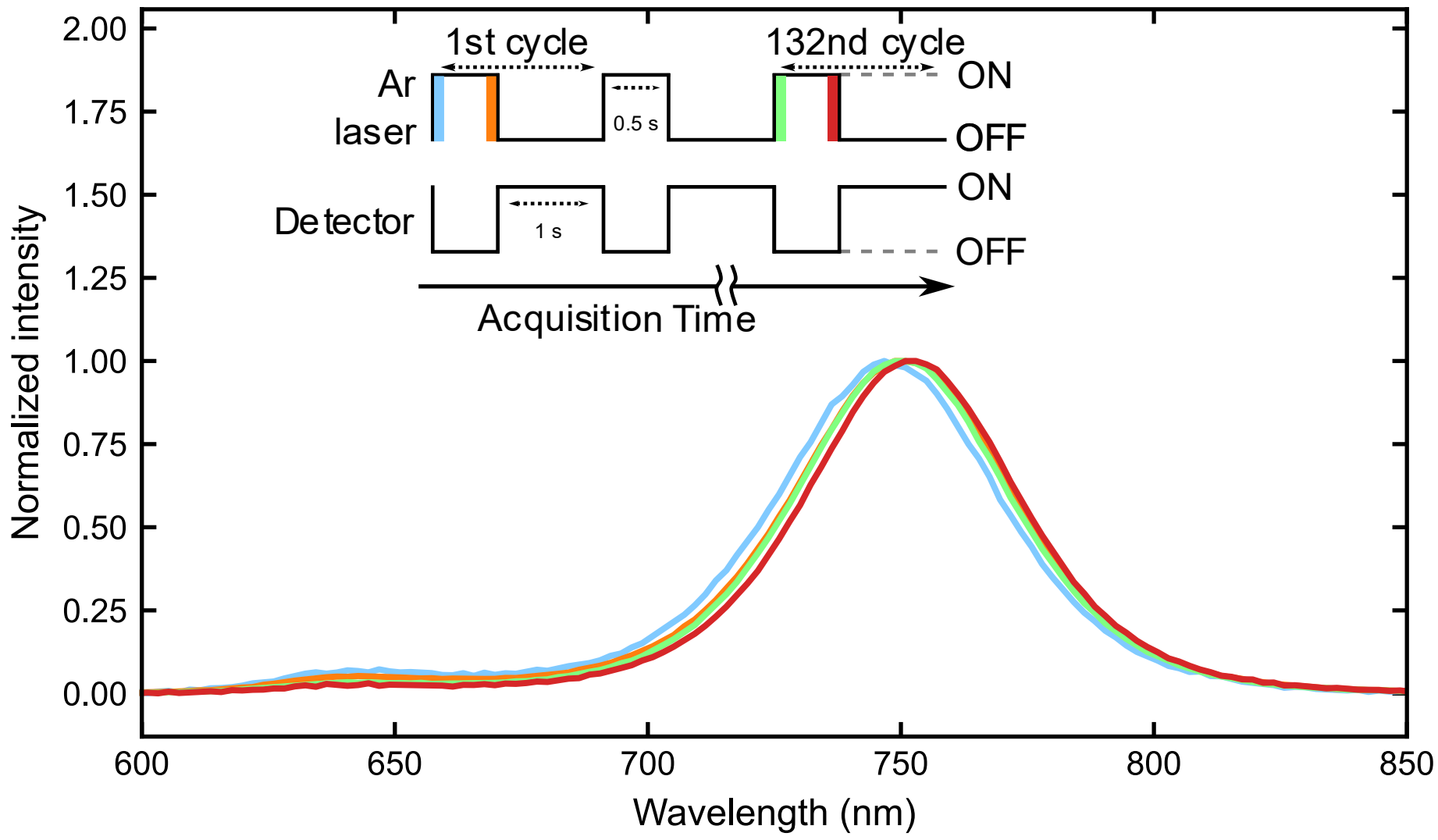

**Figure S6.** PL spectra recorded during the RWR measurements of the segregated state using the alternating CW Ar-ion laser illumination and TRPL detection (Figure S5). The inset illustrates the measurement sequence, in which the CW laser and TRPL detector are operated alternately during the acquisition; the spectra shown correspond to measurements at different points during the acquisition, including the first and 132$^{nd}$ cycles. The PL spectra exhibit only a small change over the measurement, indicating that the sample remained predominantly in the segregated state throughout the TRPL acquisition.

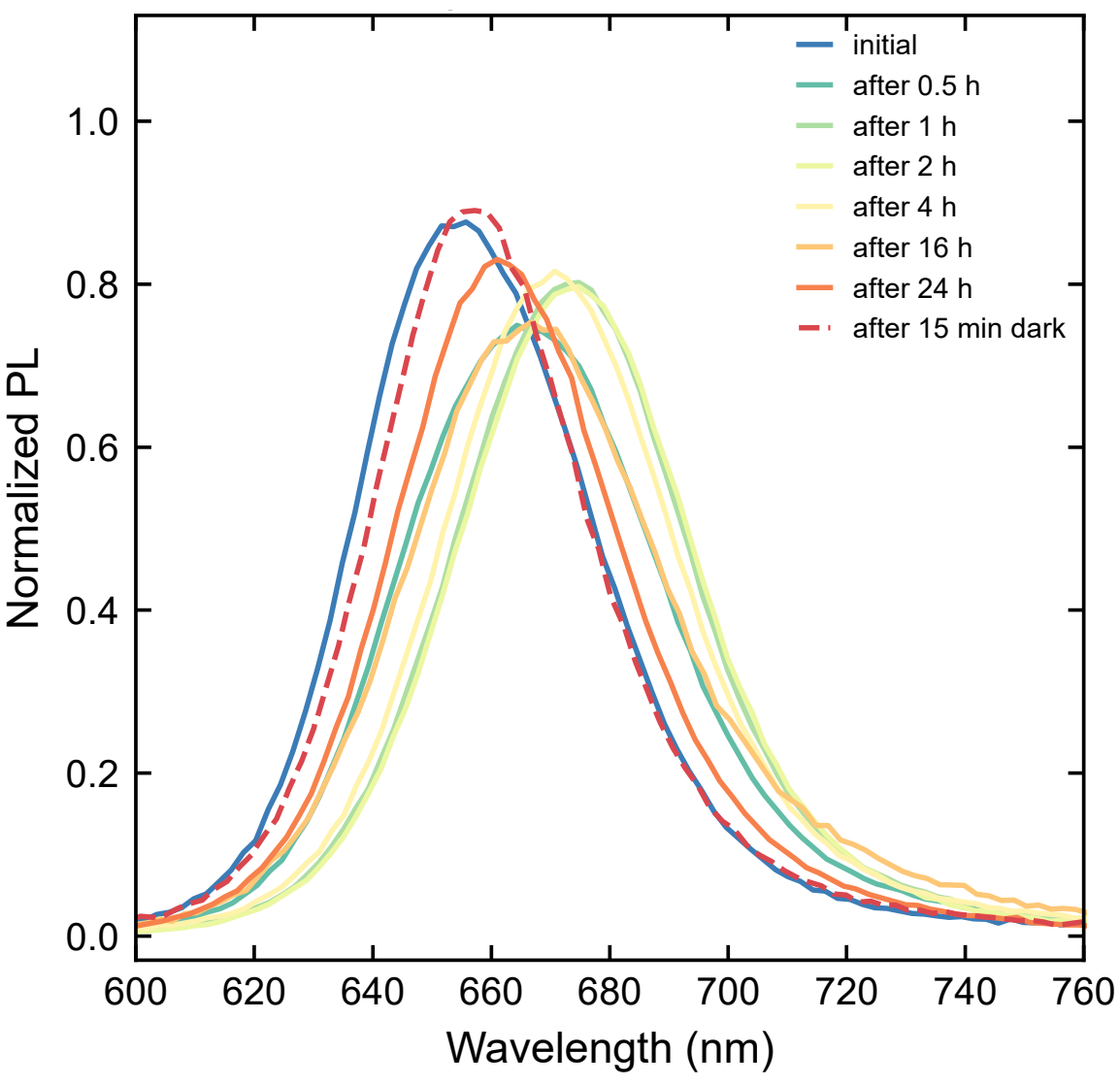

**Figure S7.** Stability of the mixed state under the condition N=100, M=10000. The spectra were taken after 0.5h, 1h, 2h, 4h, 16h and 24h from the start of excitation. After 24 hours, the sample was left in dark for 15 minutes, and then the spectrum was measured again. The measurements were performed at 2 $\mu$J cm$^{-2}$ fluence.

### Supplementary Note S6: Stability of the mixed state